\documentclass[sigconf,screen]{acmart}

\AtBeginDocument{%
  }

\usepackage[normalem]{ulem}
\newcommand{\smallsection}[1]{\smallskip\noindent\textbf{#1}}
\usepackage{soul}
\usepackage{dirtytalk}
\usepackage{subcaption}
\usepackage{caption}
\usepackage{placeins}
\usepackage{rotating}

\begin{document}

\title{
“Detective Work We Shouldn’t Have to Do”: Practitioner Challenges in Regulatory-Aligned Data Quality in Machine Learning Systems}

\author{Yichun Wang}
\orcid{1234-5678-9012}
\affiliation{%
  \institution{University of Amsterdam}
  \country{The Netherlands}}

\author{Kristina Irion}
\affiliation{%
 \institution{University of Amsterdam}
 \country{The Netherlands}}

\author{Paul Groth}
\affiliation{%
  \institution{University of Amsterdam}
  \country{The Netherlands}
}

\author{Hazar Harmouch}
\affiliation{%
 \institution{University of Amsterdam}
 \country{The Netherlands}}



\begin{abstract}
Ensuring data quality in machine learning (ML) systems has become increasingly complex as regulatory requirements expand. In the European Union (EU), frameworks such as the General Data Protection Regulation (GDPR) and the Artificial Intelligence Act (AI Act) articulate data quality requirements that closely parallel technical concerns in ML practice, while also extending to legal obligations related to accountability, risk management, and human rights protection. This paper presents a qualitative interview study with EU-based data practitioners working on ML systems in regulated contexts. Through semi-structured interviews, we investigate how practitioners interpret regulatory-aligned data quality, the challenges they encounter, and the supports they identify as necessary. Our findings reveal persistent gaps between legal principles and engineering workflows, fragmentation across data pipelines, limitations of existing tools, unclear responsibility boundaries between technical and legal teams, and a tendency toward reactive, audit-driven quality practices. We also identify practitioners’ needs for compliance-aware tooling, clearer governance structures, and cultural shifts toward proactive data governance.
\end{abstract}



\received{20 February 2007}
\received[revised]{12 March 2009}
\received[accepted]{5 June 2009}

\maketitle

\section{Introduction}

Data quality plays a central role in the reliability, fairness, and safety of ML systems~\cite{Holstein2018ImprovingFI, sambasivan2021everyone, Gebru2018DatasheetsFD}. Poor-quality data can lead to biased outcomes, unstable models, and system failures that affect individuals and organisations alike~\cite{QuioneroCandela2009DatasetSI, sculley2015hidden, northcutt2021confident}. In much of the technical research, data quality in ML is framed in terms of performance and robustness. Practitioners focus on avoiding label noise, distribution shift, data drift, and inconsistencies that degrade model accuracy or stability~\cite{Patrini2016MakingDN, Breck2017TheMT,Ovadia2019CanYT, Dong2019TowardsTD}. To this end, they deploy validation checks, monitoring, and pipeline orchestration tooling to keep pipelines stable in production~\cite{Schelter2018AutomatingLD, Zaharia2018AcceleratingTM}.

Beyond these technical considerations, in the EU, data quality is also a regulatory requirement tied to accountability, risk management, and the protection of fundamental rights. EU regulations such as the General Data Protection Regulation (GDPR) and the Artificial Intelligence Act (AI Act) introduce explicit requirements related to data quality~\cite{GDPR2016, AIA2024}. The GDPR establishes principles such as accuracy, data minimisation, purpose limitation, and accountability in relation to the processing of personal data~\cite{hoofnagle2019european, goddard2017eu}. The AI Act further requires appropriate and representative training, validation, and testing data, bias mitigation measures, documentation, and ongoing risk management for high-risk AI systems~\cite{novelli2024taking, laux2023trustworthy, Veale2021DemystifyingTD}. Together, these regulations require data quality practices that satisfy both technical and legal demands.

In practice, aligning data quality with regulatory requirements is not straightforward. Modern ML systems rely on complex pipelines that span data ingestion, transformation, feature engineering, training, deployment, and monitoring~\cite{rangineni2023analysis, priestley2023survey, sculley2015hidden, Breck2017TheMT}. Responsibilities for data quality are distributed across teams, tools, and organisational boundaries~\cite{sambasivan2021everyone, Passi2019ProblemFA, Madaio2020CoDesigningCT,10.1145/3715275.3732017}. Legal requirements are expressed as high-level principles, while engineers work with concrete constraints, metrics, and delivery timelines~\cite{Edwards2017SlaveTT, Basin2018OnPA, Passi2019ProblemFA}. This structural mismatch complicates the operationalisation of regulatory requirements into concrete data quality practice across multiple pipeline components and organisational contexts. To ground this challenge, we conceptualise "regulatory-aligned data quality" as the degree to which technical data quality practices both meet engineering standards and satisfy applicable legal obligations. We systematically map commonly used data quality dimensions to the corresponding requirements in the GDPR and the AI Act (see Appendix~\ref{sec:appendix-dq} Table~\ref{tab:app-data-quality}). However, how practitioners enact this alignment in practice remains underexplored. As a result, practitioners often face uncertainty about how regulatory requirements should be interpreted and operationalised in everyday data quality assurance practices~\cite{raji2020closing, veale2018fairness, labadie2023building, sirur2018we}. There is limited empirical insight into how they navigate this space, particularly in the EU context.

This paper addresses this gap by examining regulatory-aligned data quality from the perspective of practitioners. Practitioners play a central role in operationalising regulatory requirements through their day-to-day data work, yet their perspectives on regulatory-aligned data quality remain underexplored. We ask the following research question:
\begin{center}
\emph{What real-world challenges do practitioners face in ensuring regulatory-aligned data quality in ML systems, and what forms of support do they need to address these challenges in practice?}


\end{center}

To effectively address this central question, the study will delve into the following specific
sub-questions:
\begin{itemize}
\item {\textbf{SRQ1:} How do practitioners currently interpret and operationalise key regulatory-aligned data quality dimensions in their day-to-day ML workflows, and what factors shape these practices?}

\item {\textbf{SRQ2:} What tools, methods, and infrastructures do practitioners use to manage regulatory-aligned data quality today, and what additional capabilities do they perceive as necessary to meet regulatory obligations?}

\item {\textbf{SRQ3:} How do collaboration patterns between technical teams and legal or compliance teams influence the implementation of regulatory-aligned data quality?}

\end{itemize}

\begin{figure*}[t]
  \centering
  \includegraphics[width=\textwidth]{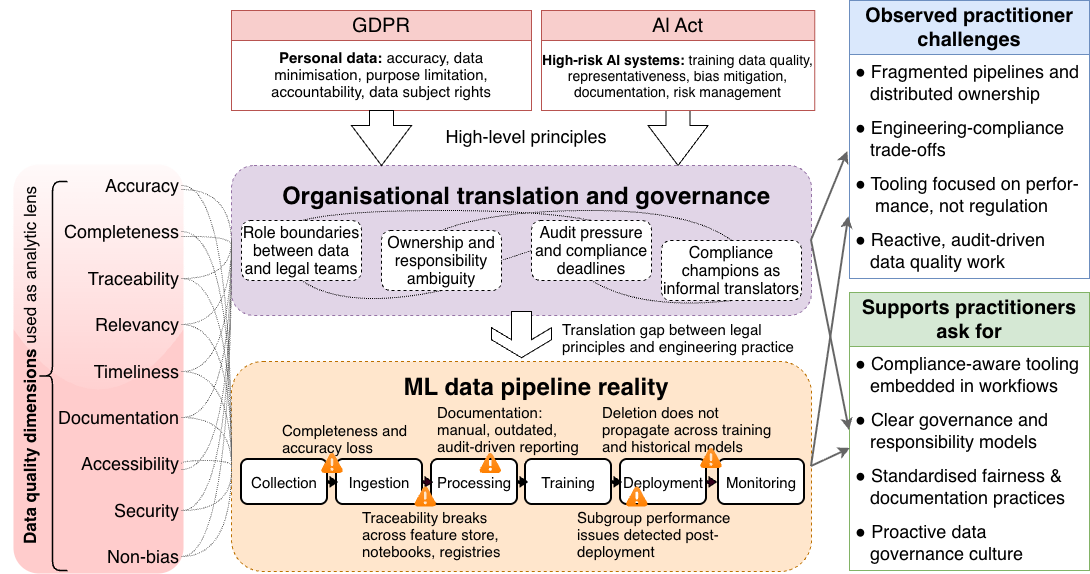}
  \caption{Regulatory-Aligned Data Quality in Practice - Challenges and Desiderata}
  \label{fig:overview}
\end{figure*}

To answer these questions, we conducted semi-structured interviews with fourteen EU-based practitioners working on ML systems that process personal data or operate in regulated domains. Participants covered a range of roles, including data engineering, data science, ML engineering, and compliance. The interviews were structured around nine data quality dimensions including accuracy, traceability, relevancy, non-bias (see Figure~\ref{fig:overview}). Participants were asked to reflect on their experience with data quality issues that arise from legal compliance requirements. We also introduced three scenario-based vignettes inspired by the GDPR and AI Act requirements (see Appendix~\ref{sec:appendix-vignettes}). This design encouraged participants to reflect on concrete practices rather than abstract principles.

This paper makes three main contributions. First, it provides an empirical account of how practitioners in the EU interpret and enact regulatory-aligned data quality across ML pipelines. Second, it identifies cross-cutting challenges that shape this work, including the translation of legal principles into engineering tasks, fragmentation of data pipelines, limitations of existing tools, and gaps in collaboration and responsibility sharing between technical and legal teams. Third, it distils practitioner-informed directions for future research on compliance-aware data quality, including opportunities for new forms of technical infrastructure, governance structures, and operational guidance. Figure~\ref{fig:overview} provides an overview of the study design and analysis outcomes.

\section{Literature review}
\label{sec:related_work}

This study sits at the intersection of data quality and EU regulation. Prior work has developed rich technical methods, governance frameworks, and legal analyses. Yet these strands often evolve separately. In this section, we review major literature from technical, regulatory, and empirical perspectives, and identify gaps that motivate our practitioner-focused study of regulatory-aligned data quality in practice.

\subsection{Data quality \& ML: the technical perspective}
Data quality has traditionally been studied in data management through dimensions such as accuracy, completeness, timeliness, consistency, relevance, and accessibility~\cite{wang1996beyond, batini2016data}. In ML, data quality is increasingly recognised as a primary determinant of model performance and robustness. Recent work emphasises data-centric AI, arguing that systematic data improvement such as curation, coverage, and label integrity often yields larger gains than model complexity~\cite{sambasivan2021everyone, zha2025data}. Studies highlight how label noise ~\cite{northcutt2021confident}, dataset shift ~\cite{QuioneroCandela2009DatasetSI}, class imbalance~\cite{Johnson2019SurveyOD}, and representativeness~\cite{Chasalow2021RepresentativenessIS} affect model behaviour and generalisation. 

Quality assurance practices in ML pipelines reflect these concerns. Teams commonly employ schema validation, rule-based checks, and statistical tests to detect anomalies and drift ~\cite{Breck2017TheMT, Schelter2018AutomatingLD}. Feature stores and data versioning systems aim to stabilise training and inference by controlling data dependencies~\cite{Fu2023FEASTAC}. Experiment tracking frameworks record configurations and metrics to support reproducibility ~\cite{Zaharia2018AcceleratingTM}. Tooling such as Great Expectations, Deequ, and similar frameworks allow teams to declare and test data quality constraints in code~\cite{schelter2019unit,Gong_Great_Expectations}. These practices are now standard in ML operations. However, they primarily prioritise improvements to predictive performance or system reliability and do not always address the requirements associated with legal compliance.

Alongside quality checks, technical governance mechanisms have emerged. Data catalogs, lineage systems, and provenance frameworks track data flows and transformations across pipelines~\cite{Buneman2001WhyAW, Cheney2009ProvenanceID, Herschel2017ASO}. Metadata management describes datasets, features, and models~\cite{Sawadogo2020OnDL}. Documentation artefacts such as datasheets for datasets and model cards aim to improve transparency and reviewability ~\cite{Gebru2018DatasheetsFD, Mitchell2018ModelCF}. Systems such as Apache Atlas, Amundsen, DataHub, and commercial data catalogs aim to provide end-to-end visibility of data assets~\cite{apacheatlas,amundsen2019, datahub2019}. MLOps platforms like MLflow, Kubeflow, and cloud-native solutions support experiment tracking, model versioning, and deployment~\cite{zaharia2018mlflow, kubeflow}. Together, these tools and practices define a technical stack for quality assurance and governance.

Yet important gaps remain. Data quality is conceptualised as properties that can be instrumented and monitored within the technical stack. It captures how data moves and how models are trained. It rarely encodes legal requirements such as purpose limitation, lawful basis, consent, or retention constraints. Deletion propagation across derived features, caches, and historical training sets is also weak~\cite{Ginart2019MakingAF, Izzo2020ApproximateDD, Chourasia2022ForgetUT}. Fairness and subgroup evaluation often rely on ad hoc scripts that are hard to reproduce ~\cite{Holstein2018ImprovingFI, Rabonato2024ASR, Madaio2020CoDesigningCT}. Documentation is frequently manual and drifts from the living pipeline, especially in fast-evolving systems~\cite{Heger2022UnderstandingML, GinerMiguelez2024UsingLL}. Fragmentation across tools creates blind spots in end-to-end oversight. The gap between technical quality and regulatory quality is therefore built into many current approaches. Our findings corroborate this body of work and provide empirical evidence of how such gaps surface in practitioners' day-to-day compliance work.

\subsection{Regulatory perspectives on data quality}
EU regulations impose explicit demands on data quality, though they do so at a high level of abstraction. The GDPR regulates the processing of datasets containing personal data through principles of accuracy, data minimisation, purpose limitation, storage limitation, integrity and confidentiality, and accountability~\cite{GDPR2016}. These principles directly affect how personal data is collected, processed, retained, and documented. Accuracy links to record correctness and rectification duties. Minimisation and purpose limitation constrain feature selection and data reuse. Storage limitation determines retention and deletion. Integrity and confidentiality drive access control and protection. Accountability requires evidence and documentation that connect choices to legal obligations and demonstrate compliance with those obligations. Data subject rights further introduce technical demands for traceability, rectification, erasure, and portability~\cite{Veale2018WhenDP}. These rights turn technical lineage into compliance evidence. They also expose weak spots in pipelines that were not designed to support such obligations.

The AI Act extends these requirements for high-risk AI systems. It mandates data governance and management practices, representative and bias-mitigated training data, risk management, documentation, and post-market monitoring~\cite{AIA2024}. Legal scholarship has analysed the normative goals of these provisions and their implications for AI accountability~\cite{Veale2021DemystifyingTD, Laux2023TrustworthyAI}. However, these duties are principle based and technology neutral~\cite{Gamito2024ArtificialIC}. It specifies targets but leave implementation choices open.

Applying legal requirements to concrete ML projects requires case-specific interpretation, which creates a persistent translation problem~\cite{binns2018fairness, veale2018fairness, Veale2021DemystifyingTD,sargeant2025formalising}. Regulations articulate principles, while practitioners need concrete actions and artefacts. Prior interdisciplinary work highlights the difficulty of operationalising legal norms in technical systems~\cite{Edwards2018EnslavingTA, Veale2018ClaritySA, Edwards2017SlaveTT, Basin2018OnPA, 10.1145/3715275.3732004}. Studies note that engineers are often left to interpret compliance obligations without sufficient guidance~\cite{Barati2021PrivacyAwareCA, Voigt2024TheEG, sirur2018we}. As our findings later show, ML practitioners expressed a clear need to know which pipeline steps, artefacts, and checks provide sufficient evidence. They need shared language across legal and technical teams.

\subsection{Empirical studies on practitioner experience and organisational dynamics}
A growing body of empirical research examines how ML practitioners engage with responsible AI practices. Interview and survey studies document challenges related to fairness, explainability, and accountability~\cite{Holstein2018ImprovingFI, Rakova2020WhereRA, madaio2024learning}. These studies show that responsibility is often diffused across teams, incentives are misaligned, and ethical practices are applied late in development cycles.

Research on governance and provenance adoption reveals similar patterns. Practitioners value traceability but struggle to maintain it across organisational boundaries and heterogeneous toolchains~\cite{sambasivan2021everyone, Roshan2024EnsembleAO}. Accountability and normative concerns are rarely embedded into everyday development workflows, instead surfacing reactively after key technical decisions have already been made~\cite{Passi2019ProblemFA}. In practice, responsibility for governance often falls to informal compliance champions who translate between legal expectations and technical workflows, but these roles are typically under-resourced and lack formal authority~\cite{Madaio2020CoDesigningCT}.

However, existing empirical work tends to focus on model-level concerns such as fairness or explainability~\cite{10.1145/3715275.3732032}. Data quality is rarely examined as a cross-cutting concern that spans ingestion, transformation, training, deployment, and monitoring. EU-specific regulatory contexts are also underrepresented. As a result, little is known about how practitioners operationalise data quality dimensions in response to concrete GDPR and AI Act demands. This study directly addresses this gap.

\begin{table*}[t]
\caption{Overview of interview respondents, including industry domain, organisational size, role, and core responsibilities.}
\label{tab:respondents}
\centering
\begin{tabular}{l p{3.2cm} p{2.2cm} p{3.6cm} p{5.4cm}}
\toprule
 & \textbf{Industry} & \textbf{Company Size} & \textbf{Role} & \textbf{Primary Responsibility} \\
\midrule
R1  & Finance (banking)            & >10,000        & Data scientist / data engineer & ML model development and monitoring for compliance \\
R2  & Finance (banking)            & >10,000        & Data scientist                 & ML model validation for compliance \\
R3  & Finance, telecommunications  & >10,000        & Data engineer                  & Data engineering (freelance) \\
R4  & Retail                       & >10,000        & ML engineer                    & Sales forecasting models \\
R5  & Finance, transportation      & >10,000        & Data scientist                 & Fraud prediction modelling \\
R6  & Transportation               & 500--1,000     & Data analyst                   & HR analytics dashboards \\
R7  & Health                       & 1,000--5,000   & Data scientist / researcher    & Disease prediction models \\
R8  & Market research              & 25--100        & Data scientist                 & Interview data analysis with LLMs \\
R9  & GenAI, transportation        & >10,000        & Data science manager           & Data science oversight and compliance \\
R10 & Finance, geography           & 1,000--5,000   & Data scientist / data collector & Property price forecasting \\
R11 & Audit                        & 1,000--5,000   & Legal and compliance           & Compliance functions \\
R12 & Audit                        & 1,000--5,000   & Legal and compliance           & Compliance functions \\
R13 & Audit                        & 1,000--5,000   & Legal and compliance           & Compliance functions \\
R14 & Audit                        & 1,000--5,000   & Legal and compliance           & Compliance functions \\
\bottomrule
\end{tabular}
\end{table*}

\subsection{Research gap}
The technical literature offers strong methods for validation, monitoring, versioning, and lineage, yet largely oriented toward performance and operational stability. Regulatory and social-legal work articulate data protection principles at a high level, but provide limited guidance on how these requirements should be operationalised in concrete ML projects. Empirical studies reveal organisational and cultural barriers. Taken together, these strands address related concerns but remain weakly connected.

This study contributes by examining how EU-based practitioners working with personal data or high-risk ML systems navigate this space in practice. Through qualitative interviews, we investigate how they interpret key data quality dimensions, which challenges they encounter when aligning their work with regulatory requirements, and what support they feel is missing. This perspective complements technical and legal approaches and provides an empirical basis for designing future tools, processes, and policies.

\section{Methodology}
\label{sec:methodology}

We adopt a qualitative, interview-based design, which is widely used in computer science research to investigate practitioners' experiences and trades-offs in real-world ML development context~\cite{kvale2009interviews, lazar2017research, Holstein2018ImprovingFI}. We interviewed fourteen practitioners based in the EU. In this study, we defined practitioners broadly as individuals involved in any stage of developing ML-based products or services. The sample included operational roles (e.g., data collectors and labellers), technical roles (e.g., data engineers, data scientists, and ML engineers), and legal or compliance roles (compliance officers and legal staff).  Table~\ref{tab:respondents} presents an overview of participant characteristics.

\subsection{Interview design}
The semi-structured interview protocol contained four main parts (see Appendix~\ref{sec:appendix-interview-outline}). First, we asked about participants’ working background and project setup, including any existing data quality standards in their daily practices. Second, we discussed a concrete case of a data quality issue that became visible due to compliance risk. Third, we introduced a set of short vignettes~\cite{deterding2011situated}, to explore how participants would act in specific regulatory or data quality scenarios. Finally, for respondents with substantial compliance experience, we asked additional questions about collaboration with legal teams, perceived challenges, trade-offs, and motivations. The protocol was adapted to each participant’s role and expertise.

Designing interviews on compliance-related topics requires care. Respondents might hesitate to speak openly due to social desirability concerns or perceived organisational risk. To reduce this pressure, we used several strategies. We explained clearly that all identifying details about individuals, organisations, products, and services would be anonymised. Before scheduling formal interviews, we held informal conversations with most participants to build trust and reduce concerns. During the interviews, we used neutral and technical language to minimise discomfort. We emphasised that the study focuses on understanding challenges and support needs, not on assessing compliance performance. We avoided any phrasing that could imply blame or wrongdoing. Participants were reminded that they could pause or stop the interview or recording at any time. This approach helped create a safe environment and encouraged open and reflective responses. The complete interview guide is available in Appendix~\ref{sec:appendix-interview-outline}.

As part of this careful design, we used vignettes as an abstract, scenario-based elicitation technique. The vignettes described hypothetical situations unrelated to participants’ own organisations, which help reduce defensiveness while prompting reflection on realistic practices and trade-offs. Prior work shows that vignettes are effective for eliciting tacit reasoning and feasibility constraints in ethically or legally sensitive domains~\cite{barter2000wanna,10.1145/2181037.2181040}. The vignettes were designed to reflect recurring regulatory tensions in the GDPR and the AI Act, including bias and subgroup performance (GDPR special categories of personal data, Art.\, 22(4); AI Act data governance, Art.\, 10(2), (3), and (5)), unclear dataset provenance (GDPR  lawful basis, Art.\,6; AI Act data governance, Art.\,10)
, and audit preparation (GDPR accountability, Art.\,5(2); AI Act documentation, Art.\,18). Participants were explicitly encouraged to describe how their teams would realistically respond, including what would not be possible or would require additional support. The full vignette descriptions are provided in Appendix~\ref{sec:appendix-vignettes}.

\subsection{Recruitment and Implementation}
We recruited participants through convenience sampling, primarily via professional networks and in-person ML and AI events, including conferences, meet-ups, and seminars. 
To ensure relevance for both the GDPR and the AI Act, we intentionally included participants whose work involved personal data (as defined by Art.~4(1) GDPR \cite{GDPR2016}) or high-risk ML systems (as listed in Annex~III AI Act \cite{AIA2024}). 
Recruitment and interviewing were conducted iteratively between April and December 2025. Data collection and analysis proceeded in parallel, allowing emerging themes to inform continued recruitment. We stopped recruiting once recurring themes stabilised in our analysis, indicating thematic saturation.

Interviews were conducted in person or via video conferencing based on participant preference. Each session lasted approximately one hour. We recorded audio for transcription. All personal and organisational identifiers were removed during transcription, and original audio files were deleted afterwards. All transcripts and notes were stored securely on local encrypted devices.

\subsection{Analysis}
We followed the iterative coding approach described by Deterding and Waters~\cite{deterding2021flexible}, combining inductive and deductive coding strategies. We began with deductive codes derived from established data quality dimensions, relevant EU regulations (the GDPR and the AI Act), and topics from our research questions. We applied these codes systematically to transcripts and field notes. Simultaneously, we remained open to emergent themes not anticipated by our initial framework. We then conducted a second round of analytic coding to identify patterns, similarities, and differences across participants. Emerging themes were refined through discussions with researchers specialising in ML, AI, and law. This iterative process enabled us to identify patterns of meaning that informed the final results.

\subsection{Limitations}
Care was taken to minimise bias throughout the study, including during interview design, recruitment, and analysis. However, several limitations should be acknowledged.

First, the recruitment process introduces sampling constraints. We aimed to include diverse domains, roles, countries, and organisational sizes, but the overall number of participants remains limited. The sample also skews toward male and larger companies. This reflects broader patterns in the AI practitioner workforce, which remains disproportionately male and concentrated in large companies~\cite{aiindex2025}. A broader sample collected over a longer period would improve representativeness. That said, the sample size aligns with qualitative research norms, for example, the average sample size in ACM CHI studies is twelve~\cite{Caine2016LocalSF}. In addition, the field of ML and data engineering evolves quickly. Participants’ understandings of emerging technologies or tools may become outdated. Nonetheless, many of the underlying needs they described, such as clearer governance and support for regulatory alignment, are likely to persist.

Second, the study relies on an interview approach, which captures perceptions rather than direct observation of organisational behaviour. Participants’ accounts may mix factual descriptions with assumptions or incomplete information shaped by their experience, knowledge level, and organisational context. Self-reporting and perception bias are therefore possible. We also did not have direct access to internal systems or organisational workflows, which limits our ability to verify or contextualise their statements. Field studies or ethnographic observations would complement this approach and provide richer evidence of how regulatory-aligned data quality is enacted in practice.

Overall, this study should be understood as an initial step toward understanding the challenges and needs of data practitioners working with regulatory requirements. It offers a snapshot of current practice and highlights areas for future research. 
\section{Interview Results}
\label{sec:results}

Participants across the study showed a broad awareness of regulatory requirements, especially those related to the GDPR. This is not surprising, as respondents were selected for having at least some experience with compliance demands. Many described the GDPR as a routine part of their work, while awareness of the AI Act was far less consistent. Knowledge also varied by industry. Participants from banking or finance reported stronger familiarity with regulatory frameworks than those in less regulated sectors. Practitioners described a mix of motivations behind their data quality and compliance practices. The majority of participants were driven by external pressures such as audits, legal obligations, and fear of fines. Participant R6 said, “GDPR fines are real. As a small company, if we get big fines, we are likely to be bankrupt, so we are extra careful.” Others were motivated by internal values, including organisational culture, leadership expectations, and personal responsibility. As respondent R5 put it, \begin{quote}“As a customer, I don’t want my data to be carelessly treated. So I become more cautious when processing users' personal data.”\end{quote} These motivations interacted with several factors that influence priorities, including the sensitivity of the data being processed, the organisation’s maturity in data governance, the availability of tools, and whether the industry is heavily regulated.

Although awareness of compliance was widespread, participants described significant uncertainty about how to translate legal principles into daily engineering tasks and how to navigate the resulting trade-offs. The following themes present the real-world challenges they reported and the kinds of support they identified as necessary.

\subsection{Real-world challenges in ensuring regulatory-aligned data quality}

\smallsection{Fragmented data pipelines and technical difficulty hindering consistent data quality practices.} Participants often emphasised that many of their regulatory-aligned data quality challenges stem directly from the fragmented way data flows through modern ML pipelines. Many described that when data moves across organisational boundaries, the teams responsible for upstream ingestion and engineering often lack the business context or domain knowledge held by the downstream modelling teams. This disconnect frequently undermines data \textit{accuracy} and \textit{completeness}. As practitioner R7
explained, upstream teams “don’t always know what actually matters for the model,” making it difficult for them to define appropriate validation constraints or recognise when important attributes are missing or incorrect. Another participant R9 summarised: “The data starts in the upstream team, and by the time it reaches us, no one really knows who validated what.” When foundational quality checks are misaligned, downstream teams inherit data they cannot fully trust but are nonetheless responsible for using in regulated ML systems.

This fragmentation also makes \textit{traceability} fragile. Several practitioners described that the data enters completed ML pipelines, they lose visibility into earlier transformations. This makes traceability “partial at best” as data passes through ingestion services, transformation pipelines, feature stores, and deployed models. Even when local logging practices exist, they rarely connect the full path from raw source to model input. This makes it difficult to demonstrate compliance with requirements for auditability or provenance. An interviewee R2 joked that if they had to recreate the full lifecycle of a dataset from source system to final model, they would have to “chase down five different teams and seven different tools.” \textit{Documentation} practices mirror this fragmentation as well. Participants described documentation as “telling different stories” across Git, notebooks, and internal wikis, with no mechanism to keep these sources aligned as data updates through the pipeline. Practitioners noted that these gaps become especially visible during audits reviews, when teams “realise everyone’s documentation tells a slightly different story,” as noted by participant R10. Fragmentation also affects \textit{accessibility} and \textit{security}, since data permissions are typically managed by infrastructure teams rather than ML teams, resulting in both over-permissive access in some areas and bottlenecks in others. 

The challenges become especially acute when regulations require personal data deletion. Participants repeatedly highlighted that deletion requests cannot be reliably propagated across an ML pipeline. Even when a record can be removed from a primary table, participants explained that traces of it persist throughout the ML lifecycle in derived features, historical datasets, and intermediate extracts. As one engineer, R5, put it, \begin{quote}“If a user asks for deletion, how do we remove them from derived features or historical model training data? Retraining becomes expensive or infeasible.”\end{quote}

Across organisations, even motivated teams struggle to uphold data quality and compliance requirements when no one holds responsibility for the full lifecycle of the data. The result is pockets of good practice, however, never the consistent, end-to-end governance that regulatory regimes assume~\cite{sambasivan2021everyone}.

\smallsection{Trade-offs between engineering goals and regulatory data quality requirements.} Participants described frequent tensions between what makes a model perform well and what makes it compliant. Through the vignettes, practitioners were prompted to reflect on how technical decisions collide with \textit{fairness, accountability}, and \textit{documentation} requirements in real settings. They helped surface how practitioners navigate these trade-offs under real constraints.

Many participants said that addressing \textbf{bias and subgroup performance} after deployment was one of the most difficult challenges. Several explained that models often perform well on aggregate metrics but hide significant drops for specific groups. They noted that fixing these gaps required more than retraining. Teams needed additional data collection, features redesign, or specialised evaluation pipelines. Yet these steps were hard to justify under strict delivery timelines. One engineer, R1, said, ~\begin{quote} “We want to fix it, but the business still wants the model live. It becomes a negotiation rather than a quality decision.”\end{quote} Participants also noted that fairness fixes can reduce model accuracy for the overall population. Several noted that fairness fixes rarely had clear acceptance criteria. This forced teams to weigh performance losses against the need to meet the AI Act requirements for monitoring and non-bias. The trade-off was rarely straightforward.

The vignette about \textbf{unclear dataset provenance} triggered strong reactions. Practitioners said they often encounter promising third-party datasets with incomplete \textit{documentation}. Many wanted to use the data because it promised performance gains, but they also recognised the regulatory risks. Several respondents said they would push back against management pressure. They would block the dataset until lawful basis and collection practices were clarified. Others admitted that in fast-paced environments, pressure to deliver sometimes wins. R4 said they might run experiments with the dataset “just to see if it helps,” even though they knew it created compliance debt that this did not meet the GDPR standards for lawful basis or the AI Act governance. Engineer R3 said, “The dataset might help the model, but if we cannot trust the source, it becomes a time bomb.” This scenario revealed a deeper pattern. Performance gains are highly valued, but provenance checks are slow and uncertain. Teams are caught between innovation incentives and accountability requirements.

The vignette on \textbf{audit preparation} highlighted further trade-offs. Participants said they could provide some evidence, such as accuracy checks, basic lineage, or examples of bias tests. Yet many openly admitted that they would struggle to produce the full documentation required under the GDPR accountability or the AI Act documentation keeping. They lacked a coherent, end-to-end narrative that linked data collection to model deployment. Participant R3 said they could only show “what we tested last month, not what happened in every step of the pipeline.” The vignette revealed a key insight. Teams know how to monitor, validate, and measure performance, but they lack systems that turn these activities into audit-ready compliance evidence.

Across vignettes, a clear pattern emerged. Practitioners understand regulatory expectations, but they operate in environments where engineering goals, time pressure, and organisational structures shape what is feasible. These trade-offs are not signs of neglect. They reflect the reality that compliance and engineering are driven by different incentives and timelines. Participants described a constant balancing act, where they try to meet legal standards while keeping systems functional and competitive.

\smallsection{Lack of tooling that supports regulatory-aligned data quality.} Participants described a growing mismatch between the tools used for day-to-day data quality management and the requirements set by regulatory frameworks. Their teams rely on a mix of commercial platforms, open-source libraries, and custom-built systems, but most of these tools target functional or performance-orientated quality. They rarely support the accountability, explainability, or auditability required under EU regulations. As a result, practitioners must fill the gaps with manual work, especially when preparing compliance documentation or risk assessments.

Participants explained that commercial tools were often too rigid to reflect domain-specific constraints, while open-source solutions offered useful components but few compliance-focused features. Many participants said they had built internal validators or monitoring layers to compensate. As engineer R3 put it, their custom tool became “the glue we needed to make sense of our data, because nothing off-the-shelf understood our domain or what compliance meant for us.” Others explained that they created internal systems simply to match the pace of product development, noting that only large organisations have the resources to do this. For smaller companies, keeping up with compliance was described as even more challenging.

The tooling deficit extended into the broader ML infrastructure. MLOps platforms such as MLflow, Azure ML, and SageMaker were valued for versioning and reproducibility, but participants criticised them for lacking features needed for regulatory-aligned data quality, such as capturing purpose, legal basis, or retention rules. Versioning tools could record \textit{what} changed but not \textit{why}, and lineage systems showed technical transformations but not their compliance implications. Several participants noted that small changes between model versions, such as adding a feature from a new data source or using a longer history of data, could introduce different legal bases, retention rules, or fairness risks. Their tools could record the technical version change but could not reveal these regulatory consequences. When their teams needed to justify why certain fields were included in a model, they often had to reconstruct the story manually. Participant R2 described this process as “detective work we shouldn’t have to do.”

These gaps were also evident in participants’ \textit{methods} and \textit{documentation} practices. Teams used a combination of statistical validation, distribution monitoring, bias testing, and human-in-the-loop review to assess data quality, but noted that these methods were designed around model performance rather than regulatory needs. Fairness assessments commonly relied on ad-hoc scripts, not standardised pipelines, and no organisation reported having unified tooling for \textit{non-bias} evaluation that would meet Art. 10 AI Act's data governance requirements.  \textit{Documentation} workflows were similarly fragmented. While practitioners referenced datasheets, model cards, and metadata tables, these materials were usually created after the fact, disconnected from pipeline execution, and quickly outdated. Respondent R2 in the financial industry described spending weeks each year producing a 200-page compliance report that translated their technical workflow into plain language. Even with the introduction of generative AI tools or internal chatbots, participants (R5 \& R6) said these assistants could summarise code or metadata but could not keep documentation automatically synchronised with real ML pipelines, forcing teams back to manual editing.

Altogether, participants’ experiences revealed a tooling ecosystem oriented towards engineering performance but not for regulatory alignment. Existing tools fall short of supporting the consistency, traceability, explainability, and documentation required for compliance. This leaves practitioners dependent on manual processes, ad-hoc scripts, domain intuition, and cross-team negotiation, which are approaches that are difficult to sustain as regulatory scrutiny increases.

\smallsection{Collaboration and responsibility gaps between data and legal teams.} Participants repeatedly emphasised that many of their challenges were not purely technical but stemmed from organisational and communication gaps between data teams and legal or compliance teams. They described considerable ambiguity in translating high-level regulatory principles from the GDPR or the AI Act into concrete, operational data quality practices. Responsibility for specific tasks was often unclear. Practitioner R6 captured this confusion by noting, ~\begin{quote}“We thought legal was supposed to review these procedures. Legal thought we were controlling it.”
\end{quote} These blurred boundaries created uncertainty and delays, especially when a system neared deployment or an audit approached.

Collaboration practices reflected these structural tensions. Participants explained that cooperation between data and legal teams was usually informal and reactive. Both sides expressed a desire to be involved earlier in the process, rather than being consulted only at the end. When compliance entered late, technical teams struggled to adjust pipeline design, feature sets, or documentation under time pressure. As practitioner R6 put it, “We hope we can be involved earlier in the compliance procedures.” Legal teams expressed similar frustrations, noting that earlier engagement would help them provide more meaningful guidance. In some organisations, technical teams relied on internal compliance champions to bridge this gap, but these individuals carried significant workloads and spent considerable time translating regulatory guidance into actionable engineering steps.

These accounts show that collaboration and responsibility boundaries between data and legal teams remain loosely defined in many organisations. Without shared language, regular communication, or clearer ownership, regulatory aligned data quality relies heavily on cross-team negotiation and individual initiative. The result is a compliance process driven by ad hoc negotiation instead of a stable, integrated workflow.

\smallsection{Data quality work is largely invisible, unrewarded, under-resourced, and reactive.} Participants described data quality work as essential but often receiving limited attention, especially when it carried regulatory implications. Many described it as invisible labour that is necessary but rarely planned, rewarded, or adequately resourced. As a result, regulatory-aligned quality practices tend to be surfaced and triggered by audits or incidents, rather than embedded proactively into ML workflow.

Several practitioners noted that their teams tried to reconstruct lineage only after something failed or when a regulator requested evidence. Engineer R3 said their team only created lineage diagrams “when something goes wrong,” and even then, the process was manual and incomplete. Participants also noted compromises in security and access control. Under delivery pressure, teams sometimes granted broad access to speed up development, even when this conflicted with internal policy. Review cycles and regular risk assessments were also rare. Some respondents said compliance was viewed as a one-time hurdle. Once a legal team approved the initial design, teams assumed they were “free to do whatever we want” until the next major review (R11). This mindset made it difficult to maintain regulatory-aligned data quality as models evolved or as new data sources were introduced.

Across interviews, participants shared a sense that regulatory-aligned data quality work becomes visible only when problems emerge. They said it required time, coordination, and efforts that were rarely recognised in performance evaluations or project timelines. Without dedicated time, resources, and recognition, regulatory-aligned data quality is often addressed reactively rather than as a routine part of ML practice.

\subsection{Supports practitioners need to manage regulatory-aligned data quality}

\smallsection{Technical supports needed.} Participants said they needed tools that could connect engineering work with regulatory requirements. Many current systems could validate data or track model versions, but they did not help teams understand legal implications. Practitioners asked for compliance-aware tooling that fits directly into ML workflows. They wanted lineage systems that record not only how data moves but also why it is allowed to move, including purpose, retention rules, and data source conditions. Several also mentioned the need for deletion and retention mechanisms that work across pipelines, rather than only at the storage layer. Participants said this would reduce the manual “detective work” they now perform when responding to audits or deletion requests. Fairness and bias tools were another gap. People wanted standardised, repeatable fairness assessments, not ad-hoc scripts. Documentation tools that stay in sync with pipelines were also seen as essential. They hoped for platforms that surface risks early and give clear signals when a dataset, feature, or model version requires legal review. Participant R1 summarised this need by saying, ~\begin{quote}“We want tools that tell us the compliance story, not just the technical one.”\end{quote}

\smallsection{Organisational and process-level supports needed.} Practitioners emphasised that technical tools alone were not enough. They needed clearer governance structures and stable processes to support regulatory-aligned data quality management. Participants called for explicit roles that assign responsibility, along with regular review cycles. They also wanted shared procedures for dataset approval, fairness evaluation, and audit preparation. Some said they wished for standard templates or checklists that would guide them through regulated steps, especially when preparing documentation or evaluating new datasets. They wanted simple explanations of what regulators expect and how these expectations connect to daily engineering tasks. Some suggested cross-functional working groups that include data teams, legal teams, and product owners. These supports would help teams understand what good compliance looks like and when to involve legal specialists.

\smallsection{Cultural supports and mindset shifts.} Participants said that meeting regulatory-aligned data quality also needs a cultural shift. Many described a mindset where data quality and compliance are only done when required rather than ongoing practices. They wanted leadership to recognise that quality work takes time and should be included in project planning. They hoped for a culture where teams address quality proactively, rather than reacting only when audits or incidents occur. Others noted that organisations should reward work that prevents issues, not only work that delivers new features. Practitioners also emphasised the need for shared language between technical and legal staff. They believed that a more open culture, where teams discuss regulatory risks early and regularly, would reduce firefighting and make governance more sustainable. A more collaborative culture, supported by early involvement from all teams, was seen as essential for sustainable and trustworthy ML systems.
\section{Discussion and Recommendations}
\label{sec:discussion}

This study examined how practitioners understand and operationalise regulatory-aligned data quality in their daily work. The findings show that practitioners are aware of compliance demands, yet struggle to translate high-level regulatory principles into concrete engineering practice. These challenges arise not only from technical limitations but also from organisational structures, workflow fragmentation, and gaps in collaboration between data and legal teams. Together, these insights help reframe data quality as a socio-technical process shaped by regulation, governance, and engineering constraints. In this section, we discuss the broader meaning of these findings and outline implications for technical practice, organisational structures, and future research.

\smallsection{Reinterpreting data quality as a socio-technical practice.}
The results indicate that data quality in regulated ML settings extends beyong traditional engineering notions, while remaining grounded in them. Practitioners often approached data quality from an engineering perspective, focusing on performance, schema validation, distribution checks, or pipeline stability. Regulations, however, frame data quality as matters of accountability, transparency, and rights protection. This mismatch creates uncertainty about how to meet compliance requirements in everyday work. For example, legal teams view accuracy as the correctness of personal records, but engineers focus on statistical patterns in transformed datasets. Similar gaps appear in the interpretation of minimisation, relevance, and non-bias. These findings suggest that regulatory-aligned data quality is not a static property of data. Instead, it is a socio-technical practice, where technical infrastructures and regulatory expectations interact. Data quality practices need to explicitly account for both technical and regulatory interpretations, rather than assuming that compliance will naturally follow from existing engineering checks.

\smallsection{The regulatory–technical translation gap.}
A central insight from this study is the persistent gap between regulatory requirements and engineering workflows. The GDPR and the AI Act articulate broad principles, yet practitioners need actionable guidance tied to concrete pipeline steps. This ambiguity was especially visible from the interview vignettes. Practitioners struggled to decide how to address subgroup performance gaps, whether to use a dataset with unclear provenance, or how to prepare evidence for a regulatory audit. Participants described situations where even minor model changes, such as adding one feature or expanding a training window, triggered new compliance obligations that their tools could not represent. 

The difficulty is not a lack of willingness to comply. Rather, they reflected a lack of structures that help teams interpret regulatory requirements in context. Practitioners described this work as slow, uncertain, and often stressful. In practice, this translation problem leads to delays, incomplete compliance, and hesitation about model updates or new data sources. This indicates a need for more systematic mechanisms that explicitly connect regulatory principles to engineering artefacts and workflow transitions. Such mechanisms could help teams reason about the compliance impact of changes before issues surface during audits or incidents.

\smallsection{Implications for technical infrastructure and ML practice.}
The findings highlight the need for technical solutions that embed compliance into ML workflows. First, existing tools support validation, monitoring, and versioning, but they rarely capture the legal metadata required to trace why data is used, how long it should be retained, or what obligations apply when models change. This gap creates significant manual work during audit requests. A compliance-aware infrastructure should associate legal metadata with data assets and model versions. Similarly, they need deletion-aware systems that can propagate removal requests across derived features and historical datasets that often contain residual traces of personal data. Participants asked for lineage tools that reflect both technical and legal dimensions. Second, documentation should be more tightly integrated into ML workflows. Automated documentation should stay aligned with evolving pipelines. Third, they also called for more robust and standardised fairness assessment pipelines. Tools that can identify and manage these dependencies would reduce the operational burden of compliance. These gaps show that data quality for regulatory purposes requires new forms of ML infrastructure that integrate legal context rather than treating compliance as an external reporting task.

\smallsection{Implications for organisational structures and governance.}
Technical improvements alone will not close the compliance gap. Our findings show that organisational structures play a crucial role in regulatory-aligned data quality. Responsibility for quality tasks was often unclear. Data teams assumed that legal teams would provide detailed rules. While legal teams expected engineers to design compliant workflows. This gap created delays, uncertainty and compliance risks. Early involvement of both sides was seen as essential, yet often missing. This stands in tension with Art. 25(1) GDPR’s requirement for data protection by design and by default, which presupposes early consideration of compliance. Clearer governance structures are needed. Formalising cross-functional review cycles, clarifying ownership for accuracy checks, documentation, and fairness evaluation, and establishing shared internal guidelines can help surface compliance issues earlier and reduce last-minute rework. Organisations may also benefit from designated compliance champions, but these roles need adequate time and structural support to be effective. Better onboarding and internal guidelines can also help. Practitioners expressed a desire for simple explanations of what regulations require and how these requirements translate into technical tasks. Without shared understanding, compliance remains inconsistent and often reactive.

\smallsection{Developing a proactive culture of data governance.}
Another key insight is the need for a cultural shift toward proactive governance. Practitioners indicated data quality and compliance work as largely invisible and unrewarded. Teams often address issues only when incidents arise or audits approach. These tasks require time and coordination and should be built into project planning, resource allocation, and performance evaluation. A shift toward proactive governance would recognise data quality as foundational to trustworthy ML development. Leadership support was seen as essential to recognise data quality as part of product reliability and user trust rather than a bureaucratic burden. A more proactive culture could reduce reliance on individual initiative and create more stable, sustainable practices for regulatory alignment.

\section{Conclusion}
\label{sec:conclusion}

This paper provides an empirical view of regulatory-aligned data quality as it is experienced by practitioners working with ML systems in the EU. Rather than treating data quality as a purely technical attribute, the findings highlight how regulatory expectations reshape everyday data work across pipelines, roles, and organisations. 

In addressing ~\textbf{SRQ1}, the study shows how practitioners interpret and operationalise key regulatory-aligned data quality dimensions in practice, and how these interpretations are shaped by regulatory demands. Looking more closely at day-to-day workflows, the findings also address ~\textbf{SRQ2}. We show how existing tools, methods, and infrastructures support performance-orientated notions of data quality. Support for regulatory obligations such as documentation, traceability, and lifecycle governance remains limited. At the same time, practitioners’ accounts highlight how these limitations are often managed through manual workarounds and ad hoc practices. Finally, in relation to \textbf{SRQ3}, the study illustrates how collaboration patterns between technical teams and legal or compliance teams influence whether regulatory-aligned data quality can be implemented effectively, and where gaps in responsibility, communication, and timing persist.

By foregrounding practitioner perspectives, this study surfaces the often unseen coordination, interpretation, and trade-off work required to align ML systems with evolving regulatory frameworks. These insights underscore the need for future research that bridges legal intent and technical implementation, and that supports practitioners through compliance-aware infrastructure, clearer governance models, and sustained cross-disciplinary collaboration.

\section*{Generative AI Statement}
The authors used ChatGPT for  grammar checking and stylistic editing.
\section*{Ethical Considerations}
This study was conducted in accordance with the ACM Code of Ethics and Professional Conduct. The study protocol received approval from the authors' institutional ethics committee. Participation was voluntary and based on informed consent. Given the sensitivity of compliance-related topics, all personal and organisational identifiers were anonymised.


\bibliographystyle{ACM-Reference-Format}
\bibliography{sample-base}

\clearpage
\appendix
\section{Data Quality Dimensions and Regulatory Mapping}
\FloatBarrier
\label{sec:appendix-dq}

Table~\ref{tab:app-data-quality} maps commonly used data quality dimensions from the data engineering literature to corresponding requirements in the GDPR and the AI Act. These dimensions provide a structured lens for analysing regulatory-aligned data quality and were used to guide the design of the interview protocol, including question prompts and scenario-based vignettes.

\begin{table*}[t]
\caption{Intersection of data quality dimensions from technical and regulatory perspectives (GDPR and AI Act).}
\label{tab:app-data-quality}
\scriptsize
\centering
\begin{tabular}
{p{0.11\linewidth}p{0.19\linewidth}p{0.32\linewidth}p{0.28\linewidth}}
\toprule
\textbf{Data Quality Dimensions} & \textbf{Technical Interpretation} & \textbf{Provisions in the GDPR as applicable} & \textbf{Provisions in the AI Act as applicable}\\
\midrule
\textbf{Accuracy} & Data is correct, reliable, and free from errors. & Requires data accuracy; erase or rectify incorrect personal data -- \textbf{Art.\,5(1)(d)}, \textbf{Recital 39}\newline Right to rectification of inaccurate personal data -- \textbf{Art.\,16} & Datasets shall be error-free -- \textbf{Art.\,10(3)}\newline Data sets should be, to the best extent possible, free of errors -- \textbf{Recital 67} \\
\midrule
\textbf{Traceability} & Changes or updates of the data should be tracked. & Security measures including logs -- \textbf{Art.\,32} & Mandatory automated logging and record-keeping for traceability -- \textbf{Art.\,12(3)(b)\&(c)}, \textbf{Art.\,19} \\

\midrule
\textbf{Relevancy} & Data is appropriate and applicable to the intended task. & Personal data shall be adequate, relevant, and limited to what is necessary in relation to the purposes -- \textbf{Art.\,5(1)(c)}, \textbf{Recital 39} & Requires assessment of the dataset's suitability -- \textbf{Art.\,10(2)(e)}\newline Datasets must be relevant to the intended purpose -- \textbf{Art.\,10(3)} \\

\midrule
\textbf{Timeliness} & Data is up-to-date and available when needed. & Personal data must be kept up to date -- \textbf{Art.\,5(1)(d)} & \- \\
\midrule
\textbf{Completeness} & All necessary components of data are present and adequately recorded. & Adequate and relevant personal data for intended purposes -- \textbf{Art.\,5(1)(c)} & Datasets must be complete for the intended purpose -- \textbf{Art.\,10(3)}\newline Completeness to avoid biased or inaccurate AI outputs -- \textbf{Recital 67} \\
\midrule
\textbf{Documentation} & Clear and comprehensive information should describe the data and its management. & Accountability principle mandates documented processing activities -- \textbf{Art.\,5(2)}, \textbf{Art.\,35}, \textbf{Recital 82}\newline Requires maintaining records of processing activities -- \textbf{Art.\,30(1), (3)\&(4)} & High-risk AI systems shall be accompanied by instructions for the input data -- \textbf{Art.\,13(3)(b)(vi)}\newline Record-keeping of input data that led to a match -- \textbf{Art.\,12}\newline The technical documentation shall contain information on the data -- \textbf{Art.\,11}, \textbf{Art.\,18}, \textbf{Annex IV}, \textbf{Annex XI} \\
\midrule
\textbf{Accessibility} & Data can be accessed and used by authorised users. & Right of access for data subjects -- \textbf{Art.\,15}\newline Data portability in structured formats (facilitating data accessibility) -- \textbf{Art.\,20} & \- \\
\midrule
\textbf{Security} & Personal data should be protected from unauthorised disclosure or misuse. & Mandates integrity and confidentiality through robust safeguards -- \textbf{Art.\,5(1)(f)}\newline Privacy by design/default embedded in systems development -- \textbf{Art.\,25}\newline Requires appropriate technical and organisational measures to ensure a level of security appropriate to the risk -- \textbf{Art.\,32} & Privacy protection throughout AI lifecycle, adhering to the GDPR principles -- \textbf{Recital 69}\newline Incorporates privacy and data protection into its risk management and ethical frameworks for AI systems -- \textbf{Recital 27}\newline AI Act does not affect personal data privacy in the GDPR -- \textbf{Art.\,2(7)} \\
\midrule
\textbf{Non-Biased} & Data is impartial and free from biases, and treats groups equitably. & Automated individual decision-making should not be based on special categories of personal data -- \textbf{Art.\,22(4)}, \textbf{Recital 71} & Data governance and management practices to detect, prevent and mitigate possible biases -- \textbf{Art.\,10(2)(f)\&(g)}\newline Datasets shall have appropriate statistical properties regarding persons or groups of persons -- \textbf{Art.\,10(3)}\newline Exceptionally process special categories of personal data for bias detection and correction -- \textbf{Art.\,10(5)}, \textbf{Recital 70}\newline Attention to bias mitigation in the data sets -- \textbf{Recital 67} \\

\bottomrule
\end{tabular}
\end{table*}

\section{Interview Protocol and Outline}
\label{sec:appendix-interview-outline}

\begin{figure*}[t]
\centering
\includegraphics[width=\textwidth]{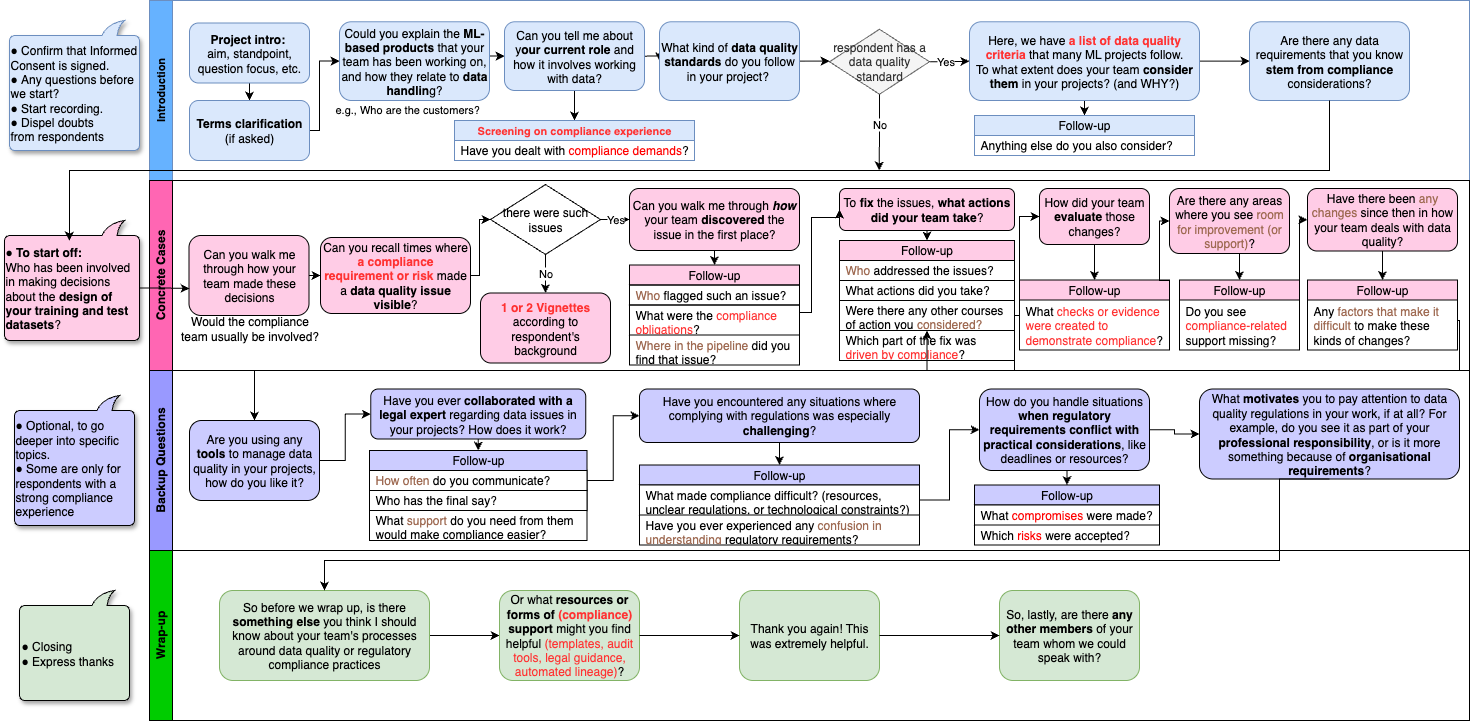}
\caption{Interview outline used in the semi-structured interviews. The outline guided discussions on data quality practices, compliance-related challenges, and regulatory-aligned trade-offs.}
\label{fig:interview-outline}
\end{figure*}

Figure~\ref{fig:interview-outline} shows the interview outline used to structure the semi-structured interviews.

\section{Interview Vignettes}
\label{sec:appendix-vignettes}

The following vignettes were used during the interviews to prompt discussion about regulatory-aligned data quality practices. The scenarios are hypothetical and intentionally abstract, and do not refer to participants’ own organisations or systems.

"Some people don’t immediately recall compliance examples, so let me give you a short scenario and ask how you’d handle it."

"Please tell us how your team would realistically handle them in your current setting. If something wouldn’t be possible, or if you’d need extra support, that’s valuable for us to hear too."

"We’re not looking for the ‘perfect’ answer, but rather the feasible choices and trade-offs your team makes in practice."

\subsection{Vignette 1: Bias and Subgroup Performance}
“Imagine you have already deployed a machine learning model. Overall performance looks fine, but monitoring shows that the model performs significantly worse for a particular subgroup, for example older users or a specific gender group.”

“How would your team realistically handle this situation in your current setting?”

“If fixing this would require trade-offs, such as retraining, collecting new data, or accepting lower overall performance, I’m interested in how you would think about those decisions.”

\subsection{Vignette 2: Unclear Dataset Provenance}
“Let me give you another scenario.”

“Your team receives a third-party dataset that could improve model accuracy, but the documentation is incomplete. It’s unclear how the data were collected, for what purpose, or under what conditions.”

“Management would like to use the dataset quickly because of delivery timelines.”

“In your current setup, how would your team realistically deal with this?”

“If something wouldn’t be possible, or if you would need extra support from legal or other teams, that’s also useful for us to understand.”

\subsection{Vignette 3: Audit Preparation and Documentation}
“Now imagine a regulator contacts your organisation.”

“They ask you to show that your training data meets quality and compliance requirements. They want evidence of accuracy checks, bias assessments, and documentation of how the data were handled.”

“What could your team realistically provide today?”

“What information would be difficult or impossible to show without additional work?”

\end{document}